\documentclass[a4paper, 10pt]{article}
\usepackage{amsmath}
\usepackage{amssymb}
\usepackage{graphicx}
\usepackage{multirow}
\usepackage{float}
\usepackage{mathtools}
\usepackage{dsfont}
\usepackage{pdflscape}
\usepackage{comment}
\usepackage{mathtools}
\usepackage{rotating}
\usepackage{graphicx}    
\usepackage{multirow}
\usepackage{placeins}
\usepackage{algorithm}
\usepackage{algorithmic}
\usepackage{setspace}

\DeclareMathOperator*{\argminA}{arg\,min}
\DeclareMathOperator*{\argmaxA}{arg\,max}
\title{Feature Selection based on the Local Lift Dependence Scale} 

\newtheorem{definition}{Definition}

\author{%
	Diego Marcondes\footnote{Universidade de S\~ao Paulo, Brazil (dmarcondes@ime.usp.br)}
	\and 
	Adilson Simonis\footnote{Universidade de S\~ao Paulo, Brazil (asimonis@ime.usp.br)}
	\and
	Junior Barrera\footnote{Universidade de S\~ao Paulo, Brazil (jb@ime.usp.br); J. Barrera was supported by grants 2013/07467 - 1 and 2015/01587 - 0, S\~ao Paulo Research Foundation (FAPESP)}}
\date{}

\begin{document}
	
	\maketitle

\begin{abstract}
This paper uses a classical approach to feature selection: minimization of a cost function applied on estimated joint distributions. However, the search space in which such minimization is performed is extended. In the original formulation, the search space is the Boolean lattice of features sets (BLFS), while, in the present formulation, it is a collection of Boolean lattices of ordered pairs (features, associated value) (CBLOP), indexed by the elements of the BLFS. In this approach, we may not only select the features that are most related to a variable $Y$, but also select the values of the features that most influence the variable or that are most prone to have a specific value of $Y$. A local formulation of Shanon’s mutual information is applied on a CBLOP to select features, namely, the \textit{Local Lift Dependence Scale}, an scale for measuring variable dependence in multiple \textit{resolutions}. The main contribution of this paper is to define and apply this local measure, which permits to analyse local properties of joint distributions that are neglected by the classical Shanon’s global measure. The proposed approach is applied to a dataset consisting of student performances on a university entrance exam, as well as on undergraduate courses. The approach is also applied to two datasets of the UCI Machine Learning Repository. 
\end{abstract}

\section{Introduction}
Let $\boldsymbol{X}$ be an m-dimensional feature vector and $Y$ a single variable. Let $\boldsymbol{\chi}$ be a feature vector, whose features are also in $\boldsymbol{X}$, and denote $\mathcal{P}(\boldsymbol{X})$ as the set of all feature vectors whose features are also in $\boldsymbol{X}$. In this scenario, we define the classical approach to feature selection, in which the search space is the Boolean lattice of features sets (BLFS).

\begin{definition}
	\label{D1}
	Given a variable $Y$, a feature vector $\boldsymbol{X}$ and a cost function $C_{Y}: \mathcal{P}(\boldsymbol{X}) \rightarrow \mathbb{R}^{+}$ calculated from the estimated joint distribution of $\boldsymbol{\chi} \in \mathcal{P}(\boldsymbol{X})$ and $Y$, the classical approach to feature selection consists in finding a subset $\boldsymbol{\chi} \in \mathcal{P}(\boldsymbol{X})$ of features such that $C_{Y}(\boldsymbol{\chi})$ is minimum.
\end{definition}
In light of Definition \ref{D1}, we note that some families of feature selection algorithms may be considered as classical approaches. In fact, according to the taxonomy of feature selection, as presented in \cite{guyon2003} for example, feature selection algorithms may be divided into three families, \textit{filters}, \textit{wrappers} and \textit{embedded} methods, being the last two classical approaches to feature selection. Indeed, in the \textit{wrappers} methods, the feature selection algorithm exists as a wrapper around a learning machine (or induction algorithm), so that a subset of features is chosen by evaluating its performance on the machine \cite{kohavi1997}, while in the \textit{embedded} methods, a subset of features is also chosen based on its performance on a learning machine, although the feature selection and the learning machine can not be separated \cite{lal2006}. Therefore, both \textit{wrappers} and \textit{embedded} methods satisfy Definition \ref{D1}, as the performance on the learning machine may be established by a cost function, so that these methods are special cases of the classical approach to feature selection. For more details about these methods see \cite{john1994,kohavi1997,hall2000,das2001,guyon2003,yu2003,lal2006}.

The main goal of the classical approach is to select the features that are most related to $Y$ according to a metric defined by the cost function. Although useful in many scenarios, this approach may not be suitable in some applications in which it is of interest to select not only the features that are most related to $Y$, but also the features values that most influence $Y$, or that are most \textit{prone} to have a specific value $y$ of $Y$. Therefore, it would be relevant to extend the search space of the classical approach to an extended space that also contemplates the range of the features, so that we may select features and subsets of their range. This extended space is a collection of Boolean lattices of ordered pairs (features,associated values) (CBLOP) indexed by the elements of the BLFS. In other words, for each $\boldsymbol{\chi} \in \mathcal{P}(\boldsymbol{X})$ we have the Boolean lattice that represents the powerset of its range $R_{\boldsymbol{\chi}}$, that is denoted by $\mathcal{P}(R_{\boldsymbol{\chi}})$, and the CBLOP is the collection of these Boolean lattices, i.e., $\{\mathcal{P}(R_{\boldsymbol{\chi}}): \boldsymbol{\chi} \in \mathcal{P}(\boldsymbol{X})\}$. If $\boldsymbol{X} = (X_{1},X_{2})$ are Boolean features, then its CBLOP is as the one in Figure \ref{FS}. Note that the circle nodes and solid lines form a BLFS, that around each circle node there is an associated Boolean lattice that represents the powerset of $R_{\boldsymbol{\chi}}$, for a $\boldsymbol{\chi} \in \mathcal{P}(\boldsymbol{X})$, and that the whole tree is a CBLOP.

A downside of this extension is that the sample size needed to perform feature selection at the extended space is greater than the one needed at the associated BLFS, what demands more refined optimal and sub-optimal algorithms in order to select the features and subsets of their range. On the other hand, the extended space brings advances to the state-of-art in feature selection, as it expands the method to a new variety of applications. As an example of such applications, we may cite market segmentation. Suppose it is of interest to segment a market according to the products that each market share is most prone to buy. Denote $Y$ as a discrete variable that represents the products sold by a company\footnote{$\mathbb{P}(Y = y)$ is the probability of an individual of the market buying the product $y \in \{1,\dots,p\}$ sold by the company.} and $\boldsymbol{X}$ as the socio-economic and demographic characteristics of the people that compose the market. In this framework, it is not enough to select the characteristics (features) that are most related to $Y$: we need to select, for each product (value of $Y$), the characteristics $\boldsymbol{\chi} \in \mathcal{P}(\boldsymbol{X})$ and their values $W \in \mathcal{P}(R_{\boldsymbol{\chi}})$ (the \textit{profile} of the people) that are prone to buy a given product, so that feature selection must be performed on a CBLOP instead of a BLFS.

We name the approach to feature selection in which the search space is a CBLOP \textit{multi-resolution}, for we may choose the features based on a global cost function calculated for each $\boldsymbol{\chi} \in \mathcal{P}(\boldsymbol{X})$ (low resolution); or choose the features and a subset of their range based on a local cost function calculated for each $\boldsymbol{\chi} \in \mathcal{P}(\boldsymbol{X})$ and $W \in \mathcal{P}(R_{\boldsymbol{\chi}})$ (medium resolution); or choose the features and a point of their range based on a local cost function calculated for each $\boldsymbol{\chi} \in \mathcal{P}(\boldsymbol{X})$ and $\boldsymbol{x} \in R_{\boldsymbol{\chi}}$ (high resolution). Formally, the \textit{multi-resolution} approach to feature selection may be defined as follows.

\begin{definition}
	\label{D2}
	Given a variable $Y$, a feature vector $\boldsymbol{X}$ and cost functions $C^{k}_{Y}: \mathcal{P}(\boldsymbol{X}) \times \mathbb{R}^{k} \rightarrow \mathbb{R}^{+}, k \in \{1,\dots,m\}$, calculated from the estimated joint distribution of $\boldsymbol{\chi} \in \mathcal{P}(\boldsymbol{X})$ and $Y$, the \textit{multi-resolution} approach to feature selection consists in finding a subset $\boldsymbol{\chi} \in \mathcal{P}(\boldsymbol{X})$ of $k \in \{1,\dots,m\}$ features and a $W \in \mathcal{P}(R_{\boldsymbol{\chi}})$ such that $C^{k}_{Y}(\boldsymbol{\chi},W)$ is minimum.
\end{definition}
The cost functions considered in this paper are local measures of dependence $C^{k}_{Y}$ such that, for each $\boldsymbol{\chi} \in \mathcal{P}(\boldsymbol{X})$ of length $k \in \{1,\dots,m\}$ and subset $W \in \mathcal{P}(R_{\boldsymbol{\chi}})$, $C^{k}_{Y}(\boldsymbol{\chi},W)$ measures the local dependence between $\boldsymbol{\chi}$ and $Y$ restricted to the subset $W$, i.e., for $\boldsymbol{\chi} \in W$. More specifically, our cost functions are based on the \textit{Local Lift Dependence Scale}, that is a scale for measuring variable dependence in multiple resolutions. In this scale we may measure variable dependence globally and locally. On the one hand, global dependence is measured by a coefficient, that summarises it. On the other hand, local dependence is measured for each subset of the feature vector range, again by a coefficient. Therefore, if the cardinality of the feature vector range is $N$, we have $2^{N} - 1$ dependence coefficients: one global and $2^{N} - 2$ local, each one measuring the influence of the feature vector in $Y$ restricted to a subset of its range. Furthermore, the \textit{Local Lift Dependence Scale} also provides a \textit{propensity} measure for each point of the joint range of the feature vector and $Y$. Note that the dependence is indeed measured in multiple resolutions: globally, for each subset of the feature vector range and pointwise.

Thus, in this paper, we extend the classical approach to feature selection in order to select not only the features, but also their values that are most related to $Y$ in some sense. In order to do so, we extend the search space of the feature selection algorithm from the BLFS to the CBLOP and use cost functions based on the \textit{Local Lift Dependence Scale} and on classical dependence measures, such as the Mutual Information, Cross Entropy and Kullback-Leibler Divergence. The feature selection algorithms proposed in this paper are applied to a dataset consisting of student performances on a university's entrance exam and on undergraduate courses in order to select exam's subjects, and the performances on them, that are most related to undergraduate courses, considering student performance on both. The method is also applied to two datasets publicly available at the UCI Machine Learning Repository \cite{lichman2013}, namely, the Congressional Voting Records and Covertype datasets. We first present the main concepts related to the \textit{Local Lift Dependence Scale}. Then, we propose feature selection algorithms based on the \textit{Local Lift Dependence Scale} and apply them to the performances and UCI datasets.

\section{\textit{Local Lift Dependence Scale}}

The \textit{Local Lift Dependence Scale }(LLDS) is a set of tools for assessing the dependence between a random variable $Y$ and a random vector $\boldsymbol{X}$ (also called feature vector). Although very simple, and consisting of well known mathematical objects, there does not seem to exist any literature that thoroughly defines and study the properties of the LLDS, even though it is highly used in marketing \cite{coppock2002} and data mining \cite[Chapter~10]{tuffery}, for example. Therefore, we present an unprecedented characterization of the LLDS, despite the fact that much of it is well known and established in the theory.

The LLDS analyses the \textit{raw} dependence between the variables, as it does not make any assumption about its kind, nor restrict itself to the study of a specific kind of dependence, e.g., linear dependence. Among LLDS tools, there are three measures of dependence, one global and two local, but with different resolutions, that assess variable dependence on multiple levels. The global measure and one of the local are based on well known dependence measures, namely, the Mutual Information and the Kullback-Leibler Divergence. In the following paragraphs we present the main concepts of the LLDS and discuss how they can be applied to the classical and \textit{multi-resolution} approaches to feature selection. The main concepts are presented for discrete random variables $\boldsymbol{X}= \{X_{1},\dots,X_{m}\}$ and $Y$ defined on $(\Omega,\mathbb{F},\mathbb{P})$, with range $R_{\boldsymbol{X},Y} = R_{\boldsymbol{X}} \times R_{Y}$, although, with simple adaptations, the continuous case follows from it.

The Mutual Information (MI), proposed by \cite{shannon}, is a classical dependence quantifier that measures the mass concentration of a joint probability function. As more concentrated the joint probability probability function is, the more dependent the random variables are and greater is their MI. In fact, the MI is a numerical index defined as
\begin{equation*}
\label{MI}
I(X,Y) \coloneqq \sum\limits_{(\boldsymbol{x},y) \in R_{\boldsymbol{X},Y}} f(\boldsymbol{x},y) \log \Bigg(\frac{f(\boldsymbol{x},y)}{g(\boldsymbol{x})h(y)}\Bigg) 
\end{equation*}
in which $f(\boldsymbol{x},y) \coloneqq \mathbb{P}(\boldsymbol{X} = \boldsymbol{x},Y = y)$, $g(\boldsymbol{x}) \coloneqq \mathbb{P}(\boldsymbol{X} = \boldsymbol{x})$ and $h(y) \coloneqq \mathbb{P}(Y = y)$ for all $(\boldsymbol{x},y) \in R_{\boldsymbol{X},Y}$. An useful property of the MI is that it may be expressed as
\begin{equation}
\small
\label{MIE}
I(\boldsymbol{X},Y) = - \sum_{y \in R_{Y}} h(y) \log h(y) + \sum\limits_{(\boldsymbol{x},y) \in R_{\boldsymbol{X},Y}} f(\boldsymbol{x},y) \log \Bigg(\frac{f(\boldsymbol{x},y)}{g(\boldsymbol{x})}\Bigg)  \coloneqq H(Y) - H(Y|\boldsymbol{X})
\end{equation} 
in which $H(Y)$ is the Entropy of $Y$ and $H(Y|\boldsymbol{X})$ is the Conditional Entropy (CE) of $Y$ given $\boldsymbol{X}$. The form of the MI in (\ref{MIE}) is useful because, if we fix the variable $Y$, and consider the features $(X_{1}, \dots, X_{m})$, we may determine which one of them is the most dependent with $Y$ by observing only the CE of $Y$ given each one of the features, as the feature that maximizes the MI is the one that minimizes the CE. In this paper, we consider the normalized MI that is given by
\begin{equation}
\label{normIM}
\eta_{\tiny \boldsymbol{X}}(Y|R_{\boldsymbol{X}}) = \frac{\sum\limits_{(\boldsymbol{x},y) \in R_{\boldsymbol{X},Y}} f(\boldsymbol{x},y) \log \Bigg(\frac{f(\boldsymbol{x},y)}{g(\boldsymbol{x})h(y)}\Bigg) }{-\sum_{y \in R_{Y}} h(y) \log h(y)} \coloneqq \frac{I(\boldsymbol{X},Y)}{H(Y)}
\end{equation}
if $H(Y) \neq 0$ and $\eta_{\tiny \boldsymbol{X}}(Y|R_{\boldsymbol{X}}) = 1$ if\footnote{If $H(Y) = 0$ then $I(X,Y) = 0$, as $0 \leq I(X,Y) \leq H(Y)$, so that $H(Y) = I(X,Y)$ and it is intuitive to define $\eta_{\tiny \boldsymbol{X}}(Y|R_{\boldsymbol{X}}) = 1$.} $H(Y) = 0$. We have that $0 \leq \eta_{\tiny \boldsymbol{X}}(Y|R_{\boldsymbol{X}}) \leq 1$, that $\eta_{\tiny \boldsymbol{X}}(Y|R_{\boldsymbol{X}}) = 0$ if, and only if, $\boldsymbol{X}$ and $Y$ are independent and that $\eta_{\tiny \boldsymbol{X}}(Y|R_{\boldsymbol{X}}) = 1$ if, and only if, there exists a function $Q: R_{\boldsymbol{X}} \rightarrow R_{Y}$ such that $\mathbb{P}(Y = Q(\boldsymbol{X})) = 1$, i.e., $Y$ is a function of $\boldsymbol{X}$.The $\eta$, MI and CE are global and general measures of dependence, that summarize to an index a variety of dependence kinds that are expressed by mass concentration.

On the other hand, we may define a LLDS local and general measure of dependence that expands the global dependence measured by the MI into local indexes and that enables local interpretation of the dependence between variables. As the MI is an index that measures the dependence between random variables by measuring the mass concentration incurred in one variable by the observation of another, it may only give evidences about the existence of a dependence, but can not assert what kind of dependence is being observed. Therefore, it is relevant to break down the MI by region, so that it can be interpreted in an useful manner and the kind of dependence outlined by it may be identified. The \textit{Lift Function} (LF) is responsible for this break down, as it may be expressed as
\begin{eqnarray*}
	L_{\tiny(\boldsymbol{X},Y)}(\boldsymbol{x},y) \coloneqq \frac{f(\boldsymbol{x},y)}{g(\boldsymbol{x})h(y)} = \frac{f(y|\boldsymbol{x})}{h(y)},  & \forall (\boldsymbol{x},y) \in R_{\boldsymbol{X},Y}
\end{eqnarray*}
in which $f(y|\boldsymbol{x}) \coloneqq \mathbb{P}(Y = y| \boldsymbol{X} = \boldsymbol{x})$. When there is no doubt about which variables the LF refers to, it is denoted simply by $L(\boldsymbol{x},y)$.

The MI is the expectation on $(\boldsymbol{X},Y)$ of the LF, so that the LF presents locally the mass concentration measured by the MI. As the LF may be written as the ratio between the conditional probability of $Y$ given $\boldsymbol{X}$ and the marginal probability of $Y$, the main interest in its behaviour is in determining for which points $(\boldsymbol{x},y) \in R_{\boldsymbol{X},Y}$ $L(\boldsymbol{x},y) > 1$ and for which $L(\boldsymbol{x},y) < 1$. If $L(\boldsymbol{x},y) > 1$ then the fact of $\boldsymbol{X}$ being equal to $\boldsymbol{x}$ increases the probability of $Y$ being equal to $y$, as the conditional probability is greater than the marginal one. Therefore, we say that the event $\{\boldsymbol{X} = \boldsymbol{x}\}$ lifts the event $\{Y = y\}$ or that instances with profile $\boldsymbol{x}$ are prone to be of the class $y$. In the same way, if $L(\boldsymbol{x},y) < 1$, we say that the event $\{\boldsymbol{X} = \boldsymbol{x}\}$ inhibits the event $\{Y = y\}$, for $f(y|\boldsymbol{x}) < h(y)$. If $L(\boldsymbol{x},y) = 1, \forall (\boldsymbol{x},y) \in R_{\boldsymbol{X},Y}$, then the random variables are independent. Note that the LF is symmetric: $\{\boldsymbol{X} = \boldsymbol{x}\}$ lifts $\{Y = y\}$ if, and only if, $\{Y = y\}$ lifts $\{\boldsymbol{X} = \boldsymbol{x}\}$. Therefore, the LF may be interpreted as $\boldsymbol{X}$ lifting $Y$ or $Y$ lifting $\boldsymbol{X}$. From now on, we interpret it as $\boldsymbol{X}$ lifting $Y$, even though it could be the other way around.

An important property of the LF is that it can not be greater than one nor lesser than one for all points\footnote{Indeed, if $L(\boldsymbol{x},y) > 1, \forall (\boldsymbol{x},y) \in R_{\boldsymbol{X},Y}$, then $f(y \mid \boldsymbol{x}) > h(y),\forall (\boldsymbol{x},y) \in R_{\boldsymbol{X},Y},$ what implies the absurd $1 = \sum_{y \in R_{Y}} f(y \mid \boldsymbol{x}) > \sum_{y \in R_{Y}} h(y) = 1$ for $\boldsymbol{x} \in R_{\boldsymbol{X}}$. With an argument analogous we see that $L(\boldsymbol{x},y)$ can not be lesser than one for all $(\boldsymbol{x},y) \in R_{\boldsymbol{X},Y}$.} $(\boldsymbol{x},y) \in R_{\boldsymbol{X},Y}$. Therefore, if there are LF values greater than one, then there must be values lesser than one, what makes it clear that the values of the LF are dependent and that the \textit{lift} is a pointwise characteristic of the joint probability function and not a global property of it. Thus, the study of the behaviour of the LF gives the full view of the dependence between the variables, without restricting it to a specific type nor making assumptions about it.

Although the LF presents a wide picture of the variables dependence, it may present it in a too high resolution, making it complex to interpret it. Therefore, instead of measuring the variables dependence for each point in the range $R_{\boldsymbol{X},Y}$, we may measure it for a \textit{window} $W \in \mathcal{P}(R_{\boldsymbol{X}})$. The dependence between $\boldsymbol{X}$ and $Y$ in the window $W$, i.e., for $\boldsymbol{X} \in W$, may be measured by the $\eta$ coefficient defined as
\begin{equation}
\label{KLD}
\eta_{\tiny \boldsymbol{X}}(Y|W) \coloneqq \frac{E\Bigg\{D_{KL}(f(\cdot|\boldsymbol{X})||h(\cdot)) \Big| \boldsymbol{X} \in W\Big\}}{E\Big\{H\big([Y|\boldsymbol{X}],Y\big) \Big| \boldsymbol{X} \in W \Big\}} = \frac{\sum\limits_{\boldsymbol{x} \in W} g(\boldsymbol{x}) \sum\limits_{y \in R_{Y}} f(y|\boldsymbol{x}) \log \frac{f(y|\boldsymbol{x})}{h(y)}}{- \sum\limits_{\boldsymbol{x} \in W} g(\boldsymbol{x}) \sum\limits_{y \in R_{Y}} f(y|\boldsymbol{x}) \log h(y)} 
\end{equation}
in which $D_{KL}(\cdot||\cdot)$ is the Kullback-Leibler divergence \cite{KL} and $H(\cdot,\cdot)$ is the cross-entropy\footnote{$H\big([Y|\boldsymbol{X}],Y\big)$ means the cross-entropy between the conditional distribution of $Y$ given $\boldsymbol{X}$ and the marginal distribution of $Y$.} \cite{CE}. The $\eta$ coefficient (\ref{KLD}) compares the conditional probability of $Y$ given $\boldsymbol{x}$, $\forall \boldsymbol{x} \in W$, with the marginal probability of $Y$, so that as greater the coefficient, as \textit{distant} the conditional probability is from the marginal one and, therefore, greater is the influence of the event $\{\boldsymbol{X} \in W\}$ in $Y$. Note that, analogously to the MI, we may write\footnote{$H\big([Y|\boldsymbol{X}]\big)$ means the Entropy of the conditional distribution of $Y$ given $\boldsymbol{X}$.}
\small
\begin{align*}
E\Big\{D_{KL}(f(\cdot|\boldsymbol{X})||h(\cdot)) \Big| \boldsymbol{X} \in W\Big\} = & \frac{- \sum\limits_{\boldsymbol{x} \in W} g(\boldsymbol{x}) \sum\limits_{y \in R_{Y}} f(y|\boldsymbol{x}) \log h(y) + \sum\limits_{\boldsymbol{x} \in W} g(\boldsymbol{x}) \sum\limits_{y \in R_{Y}} f(y|\boldsymbol{x}) \log f(y|\boldsymbol{x})}{\mathbb{P}(\boldsymbol{X} \in W)} \\
= & E\Big\{H\big([Y|\boldsymbol{X}],Y\big) \Big| \boldsymbol{X} \in W \Big\} - E\Big\{H\big([Y|\boldsymbol{X}]\big) \Big| \boldsymbol{X} \in W \Big\}
\end{align*}
\normalsize
and we have that $0 \leq \eta$\tiny$_{\boldsymbol{X}}$\normalsize$(Y|W) \leq 1$, that $\eta$\tiny$_{\boldsymbol{X}}$\normalsize$(Y|W) = 0$ if, and only if, $h(y) \equiv f(y|\boldsymbol{x}), \forall \boldsymbol{x} \in W$, and that $\eta$\tiny$_{\boldsymbol{X}}$\normalsize$(Y|W) = 1$ if, and only if, there exists a function $Q: W \rightarrow R_{Y}$ such that $\mathbb{P}(Y = Q(\boldsymbol{X})|\boldsymbol{X} \in W) = 1$. Observe that the $\eta$ coefficient of a window is also a local dependence quantifier, although its resolution is lower than that of the LF if the cardinality of $W$ is greater than one. Also note that the $\eta$ coefficient (\ref{KLD}) is an extension of (\ref{normIM}) to all subsets (windows) of $R_{\boldsymbol{X}}$, as $W = R_{\boldsymbol{X}}$ is a window. 

The three dependence coefficients presented, when analysed collectively, measure variable dependence in all kinds of resolutions: since the low resolution of the MI, through the middle resolutions of the windows $W$, until the high resolution of the LF. Indeed, the $\eta$ coefficients and the LF define a dependence scale in $R_{\boldsymbol{X}}$, that we call LLDS, that gives a dependence measure for each subset $W \in \mathcal{P}(R_{\boldsymbol{X}})$. This scale may be useful for various purposes and we outline some of them in the following paragraphs.

\subsection{Potential applications of the \textit{Local Lift Dependence Scale}}

The LLDS, more specifically the LF, is relevant in frameworks in which we want to choose a set of elements, e.g, people, in order to apply some kind of \textit{treatment} to them, obtaining some kind of \textit{response} $Y$, and are interested in maximizing the number of elements with a given response $y \in R_{Y}$. In this scenario, given the features $\boldsymbol{X}$, the LF provides the set of elements that must be chosen, that is the set whose elements have profile $\boldsymbol{x} \in R_{\boldsymbol{X}}$ such that $L(\boldsymbol{x},y)$ is greatest. Formally, we must choose the elements whose profile is
\begin{equation*}
\boldsymbol{x}_{opt}(y) = \arg \max\limits_{\boldsymbol{x} \in R_{\boldsymbol{X}}} L(\boldsymbol{x},y).
\end{equation*}
Indeed, if we choose $n$ elements randomly from our population, we expect that $n \times \mathbb{P}(Y = y)$ of them will have the desired response. However, if we choose $n$ elements from the population of all elements with profile $\boldsymbol{x}_{opt}(y) \in R_{\boldsymbol{X}}$, then we expect that $n \times \mathbb{P}(Y = y|\boldsymbol{X} = \boldsymbol{x}_{opt}(y))$ of them will have the desired response, what is $[1 - L(\boldsymbol{x}_{opt}(y),y)] \times n$ more elements when comparing with the whole population sampling framework. Observe that this framework is the exact opposite of the classification problem. In the classification problem, we want to classify an instance given its profile $\boldsymbol{x} \in R_{\boldsymbol{X}}$ into a class $y \in R_{Y}$, that may be, for example, the class $y$ such that $f(y|\boldsymbol{x})$ is maximum. On the other hand, in this framework, we are interested in, given a $y \in R_{Y}$, finding the profile $\boldsymbol{x}_{opt}(y) \in R_{\boldsymbol{X}}$ such that $f(y|\boldsymbol{x}_{opt}(y))$ is maximum. In the applications section we further discuss the differences between this framework and the classification problem, and how the LLDS may be applied to both.

Furthermore, the $\eta$ coefficient is relevant in scenarios in which we want to understand the influence of $\boldsymbol{X}$ in $Y$ by region, i.e., for each subset of $R_{\boldsymbol{X}}$. As an example of such framework, consider an image in the grayscale, in which $\boldsymbol{X} = (X_{1},X_{2})$ represents the pixels of the image and $Y$ is the random variable whose distribution is the distribution of the colors in the picture, i.e., $\mathbb{P}(Y = y) = \frac{n_{y}}{n}$ in which $n_{y}$ is the number of pixels whose color is $y \in \{1,\dots,255\}$ and $n$ is the total number of pixels in the image. If we define the distribution of $Y|\boldsymbol{X} = (x_{1},x_{2})$ properly for all $(x_{1},x_{2}) \in R_{\boldsymbol{X}}$, we may calculate $\eta_{\tiny \boldsymbol{X}}(Y|W), W \in \mathcal{P}(R_{\boldsymbol{X}})$, in order to determine the regions that are a representation of the whole picture, i.e., whose color distribution is the same of the whole image, and the regions $W$ whose color distribution differs from that of the whole image. The $\eta$ coefficient may be useful for identifying textures and recognizing patterns in images.

Lastly, the LLDS may be used for feature selection, when we are not only interested in selecting the features $\boldsymbol{\chi} \in \mathcal{P}(\boldsymbol{X})$ that are most related to $Y$, but also want to determine the features $\boldsymbol{\chi} \in \mathcal{P}(\boldsymbol{X})$ whose \textit{levels} $W \in \mathcal{P}(R_{\boldsymbol{\chi}})$ most influence $Y$. In the same manner, we may want to select the features $\boldsymbol{\chi}$ whose level $\boldsymbol{x}_{opt}(y) \in R_{\boldsymbol{\chi}}$ maximizes $L$\tiny$_{(\boldsymbol{\chi},Y)}$\normalsize$(\boldsymbol{x}_{opt}(y),y)$, for a given $y \in R_{Y}$, so that we may sample from the population of elements with profile $\boldsymbol{x}_{opt}(y) \in R_{\boldsymbol{\chi}}$ in order to maximize the number of elements of the class $y$. Feature selection based on the LLDS is a special case of the classical and multi-resolution approaches to feature selection as presented next. 

\section{Feature Selection Algorithms based on the \textit{Local Lift Dependence Scale}}

In this section we present the characteristics of feature selection algorithms based on the LLDS. We first outline the special case of the classical approach to feature selection that is based on the LLDS, and then propose \textit{multi-resolution} feature selection algorithms that are also based on the LLDS.

\subsection{Classical Feature Selection Algorithm}

Let $Y$ and $\boldsymbol{X} = (X_{1}, \dots, X_{m})$ be random variables. We call the random variables in $\boldsymbol{X}$ features and note that $\mathcal{P}(\boldsymbol{X})$, the set of all feature vectors whose features are also in $\boldsymbol{X}$, may be seen as a BLFS, in which each vector represents a subset of features. In this scheme, feature selection is given by the minimization, in the BLFS, of a cost function applied on the estimated joint probability of a feature vector and $Y$. In fact, the subset of features selected by this approach is given by
\begin{equation*}
\boldsymbol{\chi} = \argminA_{\boldsymbol{\chi}^{*} \in \mathcal{P}(\boldsymbol{X})} C_{Y}(\boldsymbol{\chi}^{*})
\end{equation*}   
in which $C_{Y}: \mathcal{P}(\boldsymbol{X}) \rightarrow \mathbb{R}^{+}$ is a cost function. The estimated error of a predictor $\Psi$ as presented in \cite[Chapter~2]{estimation}, for example, is a classical cost function. Another classical cost function is the CE as defined in (\ref{MIE}). A pseudo-code for such algorithm is presented in Algorithm \ref{A1}. The Algorithm \ref{A1} is \textit{naive}, performs an exhaustive search on the BLFS and is known to be NP-hard \cite{amaldi1998}. However, some other algorithms may be applied to find a sub-optimal solution to this problem, as sequential selection algorithms and floating search methods \cite{marill1963,whitney1971,pudil1994,somol1999,somol2006,nakariyakul2009}, or the search space may be restricted to a subspace of $\mathcal{P}(\boldsymbol{X})$. Nevertheless, there are algorithms, as the \textit{branch-and-bound} \cite{narendra1977} and the \textit{u-curve} \cite{u-curve1,u-curve2,u-curve3}, that does not perform an exhaustive search, but ensure that the selected subset of features is optimal.

{\centering
	\begin{minipage}{0.7\linewidth}
		\begin{algorithm}[H]
			\caption{Select $\boldsymbol{\chi} \in \mathcal{P}(\boldsymbol{X})$ that minimizes $C_{Y}(\boldsymbol{\chi})$.}
			\label{A1}
			\begin{spacing}{1.3}
				\begin{algorithmic}[1]
					\ENSURE $c = \infty$
					\ENSURE $\boldsymbol{\chi} = \emptyset$
					\FOR{$\boldsymbol{\chi}^{*} \in \mathcal{P}(\boldsymbol{X})$}
					\IF{$C_{Y}(\boldsymbol{\chi}^{*}) < c$}
					\STATE $c \leftarrow C_{Y}(\boldsymbol{\chi}^{*})$ 
					\STATE $\boldsymbol{\chi} \leftarrow \boldsymbol{\chi}^{*}$
					\ENDIF
					\ENDFOR
					\RETURN{$\boldsymbol{\chi}$}
				\end{algorithmic}
			\end{spacing}
		\end{algorithm}
	\end{minipage}
	\par}

\begin{sidewaysfigure}
	\centering
	\resizebox{\linewidth}{!}{%
		\includegraphics[scale = 1]{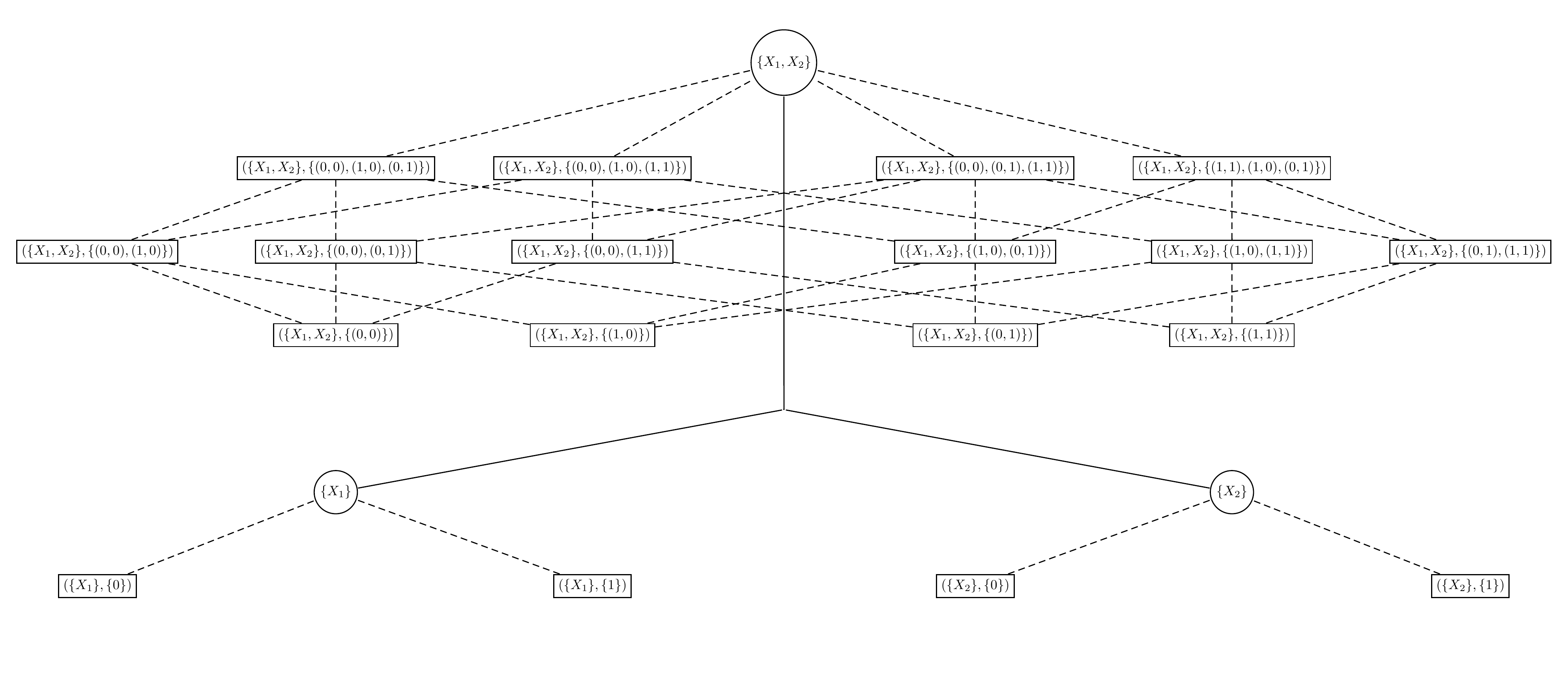}
	}
	\caption{Example of multi-resolution tree (CBLOP) for feature selection. The circle nodes and solid lines form a BLFS. The rectangular nodes and dashed lines around each circle node form a Boolean lattice. The whole tree is a CBLOP.}
	\label{FS}
\end{sidewaysfigure}

As an example of the classical approach to feature selection, suppose that $\boldsymbol{X} = (X_{1},X_{2})$, in which $X_{1}$ and $X_{2}$ are Boolean features. Then, the search space $\mathcal{P}(\boldsymbol{X})$ may be represented by a tree, i.e., a BLFS, as the one displayed in Figure \ref{FS}, considering only the circle nodes and solid lines. Algorithm \ref{A1} may be performed by walking through this tree seeking the minimum of $C_{Y}$.

\subsection{Multi-resolution Feature Selection based on the Local Lift Dependence Scale}

Feature selection based on the LLDS may be performed in three distinct resolutions. As a low resolution approach, we may select the features $\boldsymbol{\chi}$ that are most globally related to $Y$, that are given by
\begin{equation}
\boldsymbol{\chi} = \argmaxA_{\boldsymbol{\chi}^{*} \in \mathcal{P}(\boldsymbol{X})} \eta_{\tiny \boldsymbol{\chi}^{*}}(Y|R_{\boldsymbol{\chi^{*}}}).
\label{etaALL}
\end{equation} 
Note that, in this resolution, the feature selection approach is the classical one, with $1/\eta$ in (\ref{normIM}) as the cost function, i.e., Algorithm \ref{A1} may be applied to determine (\ref{etaALL}) taking $C_{Y}$ as $1/\eta$. The use of the MI as a cost function in classical feature selection algorithms is quite common in the literature (see \cite{peng2005,smieja2016,kwak2002} for example) and is not original of this paper. The search space of (\ref{etaALL}) may be restricted, sub-optimal algorithms may be applied or the discretization of the continuous features may be performed jointly, so that the subset selected in (\ref{etaALL}) is not always the subset of all features. In the applications section we show how the continuous features may be discretized jointly.

Increasing the resolution, we may be interested in finding not the features most related to $Y$, but the features \textit{levels} that most influence $Y$. In this approach the selected features and their levels are
\begin{equation}
(\boldsymbol{\chi},W) = \argmaxA_{\substack{\boldsymbol{\chi}^{*} \in \mathcal{P}(\boldsymbol{X}) \\ W^{*} \in \mathcal{P}(R_{\boldsymbol{\chi}^{*}})}} \eta_{\tiny \boldsymbol{\chi^{*}}}(Y|W^{*}).
\label{etaW}
\end{equation}
A pseudo-code for this feature selection approach is presented in Algorithm \ref{A2}. Note that the space in which the exhaustive search is conducted in Algorithm \ref{A2}, i.e., the CBLOP, is even greater than the one in Algorithm \ref{A1}. However, optimal algorithms that do not exhaustively search the space, and sub-optimal algorithms, may also be applied in this scenario, saving some computational time. Note that this approach is not suitable for the case in which the features in $\boldsymbol{X}$ are continuous, as $R_{\boldsymbol{\chi}}$, $\boldsymbol{\chi} \in \mathcal{P}(\boldsymbol{X})$, is uncountable, although the continuous features may be discretized allowing the application of the algorithm. Furthermore, as is further discussed in the applications section, this algorithm is subjected to \textit{overfitting} if the sample size is not relatively great, \footnote{As is the majority of statistical models and feature selection algorithms.} so that it may be needed to restrict its search space to a subset of the CBLOP.

{\centering
	\begin{minipage}{\linewidth}
		\begin{algorithm}[H]
			\caption{Select $\boldsymbol{\chi} \in \mathcal{P}(\boldsymbol{X})$ and $W \in \mathcal{P}(R_{\boldsymbol{\chi}})$ that maximizes $\eta$\tiny$_{\boldsymbol{\chi}}$\normalsize$(Y|W)$.}
			\label{A2}
			\begin{spacing}{1.3}
				\begin{algorithmic}[1]
					\ENSURE $c = 0$
					\ENSURE $\boldsymbol{\chi} = \emptyset$
					\ENSURE $W = \emptyset$
					\FOR{$\boldsymbol{\chi}^{*} \in \mathcal{P}(\boldsymbol{X})$}
					\FOR{$W^{*} \in \mathcal{P}(R_{\boldsymbol{\chi}^{*}})$}
					\IF{$\eta$\tiny$_{\boldsymbol{\chi}^{*}}$\normalsize$(Y|W^{*}) > c$}
					\STATE $c \leftarrow \eta$\tiny$_{\boldsymbol{\chi}^{*}}$\normalsize$(Y|W^{*})$ 
					\STATE $\boldsymbol{\chi} \leftarrow \boldsymbol{\chi}^{*}$
					\STATE $W \leftarrow W^{*}$
					\ENDIF
					\ENDFOR
					\ENDFOR
					\RETURN{$(\boldsymbol{\chi},W)$}
				\end{algorithmic}
			\end{spacing}
		\end{algorithm}
	\end{minipage}
	\par}

Finally, as a higher resolution approach, we may fix an $y \in R_{Y}$ and then look for the features levels that maximize the LF at the point $y$. Formally, the selected features and levels are given by
\begin{equation}
(\boldsymbol{\chi},\boldsymbol{x}_{opt}(y)) = \argmaxA_{\substack{\boldsymbol{\chi}^{*} \in \mathcal{P}(\boldsymbol{X}) \\ \boldsymbol{x}^{*} \in R_{\boldsymbol{\chi} ^{*}}}} L_{\tiny (\boldsymbol{\chi{*}},Y)}(\boldsymbol{x}^{*},y).
\label{Lmax}
\end{equation}
A pseudo-code to perform (\ref{Lmax}) is presented in Algorithm \ref{A3}, that is analogous to Algorithm \ref{A2}. Note that the search space of Algorithm \ref{A3} is greater than that of Algorithm \ref{A1} and smaller than that of Algorithm \ref{A2}. Nevertheless, it has the same general characteristics of Algorithm \ref{A2}: optimal algorithms that do not search all the space, and sub-optimal algorithms, may be applied; it can not be applied to continuous features; and it is subjected to \textit{overfitting}.

{\centering
	\begin{minipage}{\linewidth}
		\begin{algorithm}[H]
			\caption{Select $\boldsymbol{\chi} \in \mathcal{P}(\boldsymbol{X})$ and $\boldsymbol{x} \in R_{\boldsymbol{\chi}}$ that maximizes $L$\tiny$_{(\boldsymbol{\chi},Y)}$\normalsize$(\boldsymbol{x},y)$ for some fixed $y \in R_{Y}$.}
			\label{A3}
			\begin{spacing}{1.3}
				\begin{algorithmic}[1]
					\ENSURE $c = 0$
					\ENSURE $\boldsymbol{\chi} = \emptyset$
					\ENSURE $\boldsymbol{x} = \emptyset$
					\ENSURE $y = y$
					\FOR{$\boldsymbol{\chi}^{*} \in \mathcal{P}(\boldsymbol{X})$}
					\FOR{$\boldsymbol{x}^{*} \in R_{\boldsymbol{\chi}^{*}}$}
					\IF{$L$\tiny$_{(\boldsymbol{\chi}^{*},Y)}$\normalsize$(\boldsymbol{x}^{*},y) > c$}
					\STATE $c \leftarrow L$\tiny$_{(\boldsymbol{\chi}^{*},Y)}$\normalsize$(\boldsymbol{x}^{*},y)$
					\STATE $\boldsymbol{\chi} \leftarrow \boldsymbol{\chi}^{*}$
					\STATE $\boldsymbol{x} \leftarrow \boldsymbol{x}^{*}$
					\ENDIF
					\ENDFOR
					\ENDFOR
					\RETURN{$(\boldsymbol{\chi},\boldsymbol{x})$}
				\end{algorithmic}
			\end{spacing}
		\end{algorithm}
	\end{minipage}
	\par}

As an example of a \textit{multi-resolution} approach to feature selection based on the LLDS, suppose again that $\boldsymbol{X} = (X_{1},X_{2})$ are Boolean features. Then, for all the proposed resolutions, the selection of the features and their levels, i.e., Algorithms \ref{A1}, \ref{A2} and \ref{A3}, may be performed by walking through the tree (CBLOP) in Figure \ref{FS}. Indeed, we may calculate the global $\eta$ at the circle nodes, the $\eta$ on all windows $W$ at the rectangular nodes and the LF at the leaves, where we may determine its maximum for a fixed value $y \in R_{Y}$. Therefore, we call a tree as the one in Figure \ref{FS} a \textit{multi-resolution} tree for feature selection, where we may apply feature selection algorithms for all the resolutions of the LLDS, i.e., Algorithms \ref{A1}, \ref{A2} and \ref{A3}.

\section{Applications}

The \textit{multi-resolution} approach proposed in the previous sections is now applied to three different datasets. First, we apply it to the performances dataset, that consists of student performances on entrance exams and undergraduate courses. Then, we apply the algorithms to two UCI Machine Learning Repository datasets: the Congressional Voting Records and Covertype datasets \cite{lichman2013}.

\subsection{Performances dataset}

A recurrent issue in universities all over the world is the framework of their recruitment process, i.e., the manner of selecting their undergraduate students. In Brazilian universities, for example, the recruitment of undergraduate students is solely based on their performance on exams that cover high school subjects, called \textit{vestibulares}, so that knowing which subjects are most related to the performance on undergraduate courses is a matter of great importance to universities admission offices, as it is important to optimize the recruitment process in order to select the students that are most likely to succeed. Therefore, is this scenario, the algorithm presented in the previous sections may be an useful tool in determining the entrance exam subjects, and the performances on them, that are most related to the performance on undergraduate courses, so that students may be selected based on their performance on these subjects.

The recruitment of students to the University of São Paulo is based on an entrance exam that consists of an essay and questions of eight subjects: Mathematics, Physics, Chemistry, Biology, History, Geography, English and Portuguese. The selection of students is entirely based on this exam, although the \textit{weights} of the subjects differ from one course to another. In the exact sciences courses, as Mathematics, Statistics, Physics, Computer Science and Engineering, for example, the subjects with greater weights are Portuguese, Mathematics and Physics, as those are the subjects that are qualitatively most related to what is taught at these courses. Although weights are given to each subject in a systematic manner, it is not known what subjects are \textit{indeed} most related to the performance on undergraduate courses. Therefore, it would be of interest to measure the relation between the performance on exam subjects and undergraduate courses and, in order to do so, we apply the algorithms proposed on the previous sections.

The dataset to be considered consist of \textit{8,353} students that enrolled in  \textit{28} courses of the University of São Paulo between 2011 and 2016. The courses are those of its Institute of Mathematics and Statistics, Institute of Physics and Polytechnic School, and are in general Mathematics, Computer Science, Statistics, Physics and Engineering courses. The variable of interest (\textit{Y}) is the weighted mean grade of the students on the courses they attended in their first year at the university (the weights being the courses credits), and is a number between zero and ten. The features, that are denoted $\boldsymbol{X} = (X_{1},X_{2},X_{3},$ $X_{4},X_{5},X_{6},X_{7},X_{8},X_{9})$, are the performances on each one of the eight entrance exam subjects, that are numbers between zero and one, and the performance on the essay, that is a number between zero and one hundred. 

In order to apply the proposed algorithm to the data at hand, it is necessary to conveniently discretize the variables and, to do so, we take into account an important characteristic of the data: the scale of the performances. The scale of the performances, both on the entrance exam and the undergraduate courses, depend on the course and the year. Indeed, the performance on the entrance exam of students of competitive courses is better, as only the students with high performance are able to enrol in these courses. In the same way, the performances differ from one year to another, as the entrance exam is not the same every year and the teachers of the first year courses also change from one year to another, what causes the scale of the grades to change. Therefore, we discretize all variables by tertiles inside each year and course, i.e., we take the tertiles considering only the students of a given course and year. Furthermore, we do not discretize each variable by itself, but rather discretize the variables jointly, by a method based on distance tertiles, as follows.

Suppose that at a step of the algorithm we want to measure the relation between \textit{Y} and the features $\boldsymbol{\chi} \in \mathcal{P}(\boldsymbol{X})$. In order to do so, we discretize \textit{Y} by the tertiles inside each course and year, e.g., a student is in the third tertile if he is on the top one third students of his class according to the weighted mean grade, and discretize the performance on $\boldsymbol{\chi}$ \textit{jointly}, i.e., by discretizing the distance between the performance of each student on these subjects and zero by its tertiles. Indeed, students whose performance is close to zero have low joint performance on the subjects $\boldsymbol{\chi}$, while those whose performance is far from zero have high joint performance on the subjects $\boldsymbol{\chi}$. Therefore, we take the distance between each performance and zero, and then discretize the distance inside each course and year, e.g., a student is at the first tertile if he is on the bottom students of his class according to the joint performance on the subjects $\boldsymbol{\chi}$. The Mahalanobis distance \cite{mahalanobis} is used, as it takes into account the variance and covariance of the performance on the subjects $\boldsymbol{\chi}$. 

\begin{figure}[ht]
	\centering
	\caption{Discretization of the joint performance on Mathematics and Physics of Statistics students that enrolled at the University of São Paulo in 2011 and 2012 by the tertiles of the Mahalanobis distance inside each year.}
	\label{F1}
	\includegraphics[width = \linewidth]{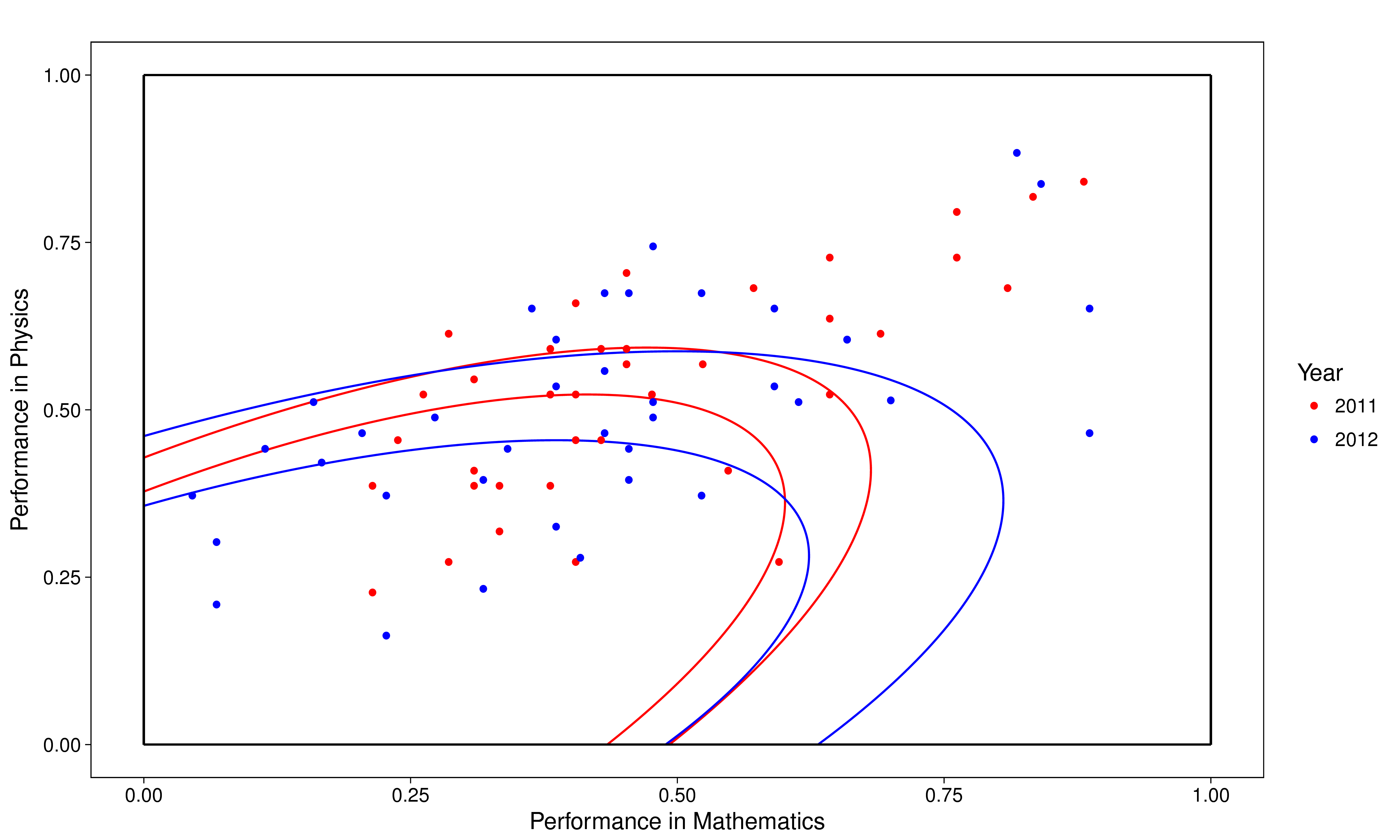}
\end{figure}

As an example, suppose that we want to measure the relation between the performances on Mathematics and Physics and the weighted mean grade of students that enrolled in the Statistics undergraduate course in 2011 and 2012. In order to do so, we discretize the weighted mean grade by year and the performance on Mathematics and Physics by the Mahalanobis distance between it and zero, also by year, as is displayed in Figure \ref{F1}. Observe that each year has its own ellipses that partition the performance on Mathematics and Physics in three and the tertile of a student depends on the ellipses of his year. The process used in Figure \ref{F1} is extended to the case in which there are more than two subjects and one course. When there is only one subject, the performance is discretized in the usual manner inside each course and year. The LF between the weighted mean grade and the joint performance on Mathematics and Physics is presented in Table \ref{T1}. From this table we may search for the maximum lift or calculate the $\eta$ coefficient for its windows. In this example, we have\footnote{$(\text{M,P}) = (\text{Mathematics,Physics})$.} $\eta$\tiny$_{(\text{M,P})}$\normalsize$(Y|R_{(\text{M,P})}) = 0.0387$. 

\begin{table}[H]
	\centering
	\caption{The Lift Function between the weighted mean grade, discretized by year, and the joint performance on Mathematics and Physics, discretized by the Mahalanobis distance inside each year, of Statistics students that enrolled at the University of São Paulo in 2011 and 2012. The numbers in parentheses represent the quantity of students in each category.}
	\label{T1}
	\begin{tabular}{ccccc}
		\hline
		Mathematics & \multicolumn{3}{c}{Weighted Mean Grade} & Relative \\ \cline{2-4}
		and Physics & Tertile 1 & Tertile 2 & Tertile 3 & Frequency  \\ 
		\hline
		Tertile 1 & 0.975 (9) & 1.46 (13) & 0.563 (5) & 0.34 \\ 
		Tertile 2 & 1.01 (9) & 0.935 (8) & 1.05 (9) & 0.33 \\ 
		Tertile 3 & 1.01 (9) & 0.584 (5) & 1.4 (12) & 0.33 \\ 
		\hline
		Relative Frequency & 0.342 & 0.329 & 0.329 & 1 \\ 
		\hline
	\end{tabular}
\end{table}

The proposed algorithm is applied to the discretized variables using three cost functions. First, we use the $\eta$ coefficient on the window that represents the whole range of the features in order to determine what are the subjects (features) that are most related to the weighted mean grade, i.e., the features (\ref{etaALL}). Then, we apply the algorithm using as cost function the $\eta$ coefficient for all windows in order to determine the subjects performances (features and window) that are most related to the weighted mean grade, i.e., the subjects and performances (\ref{etaW}). Finally, we determine what are the subjects and their performance that most lift the weighted mean grade third tertile, i.e., the subjects and performances (\ref{Lmax}) with $y = \text{Tertile 3}$. 

The subjects that are most related to the weighted mean grade, according to the proposed discretization process and the $\eta$ coefficient (\ref{normIM}), are\footnote{$(\text{M,P,C,B,Po}) = (\text{Mathematics,Physiscs,Chemistry,Biology,Portuguese})$.} $\boldsymbol{\chi} = (\text{M,P,C,B,Po})$ and $\eta$\tiny$_{\boldsymbol{\chi}}$\normalsize$(Y|R_{\boldsymbol{\chi}}) = 0.0354$. The LF between the weighted mean grade and $\boldsymbol{\chi}$ is presented in Table \ref{T2}. The features $\boldsymbol{\chi}$ are the ones that are in general most related to the weighted mean grade, i.e., are the output of the classical feature selection algorithm that employs the inverse of the global $\eta$ coefficient as the cost function (Algorithm \ref{A1}). Therefore, the recruitment of students could be optimized by taking into account only the subjects $\boldsymbol{\chi}$.

\begin{table}[H]
	\centering
	\caption{The Lift Function between the weighted mean grade and the joint performance on $\boldsymbol{\chi} = (\text{Mathematics, Physiscs, Chemistry, Biology, Portuguese})$. The numbers in parentheses represent the quantity of students in each category.}
	\label{T2}
	\begin{tabular}{ccccc}
		\hline
		Performance & \multicolumn{3}{c}{Weighted Mean Grade} & Relative \\ \cline{2-4}
		in $\boldsymbol{\chi}$ & Tertile 1 & Tertile 2 & Tertile 3 & Frequency \\ 
		\hline
		Tertile 1 & 1.33 (1,277) & 1.1 (1,018) & 0.566 (533) & 0.34 \\ 
		Tertile 2 & 0.992 (921) & 1.06 (951) & 0.954 (871) & 0.33 \\ 
		Tertile 3 & 0.669 (630) & 0.848 (775) & 1.49 (1,377) & 0.33 \\ 
		\hline
		Relative Frequency & 0.339 & 0.329 & 0.333 & 1 \\ 
		\hline
	\end{tabular}
\end{table}

Applying Algorithms \ref{A2} and \ref{A3} we obtain the same result, that the performance, i.e., window, that is most related to the weighted mean grade and that most \textit{lifts} the third tertile of the weighted mean grade is the third tertile in Mathematics, for which\footnote{M = Mathematics.} $\eta$\tiny$_{M}$\normalsize$(Y|\{\text{Tertile 3}\}) = 0.0575$ and $L$\tiny$_{(M,Y)}$\normalsize$(\text{Tertile 3,Tertile 3}) = 1.51$. The LF between the weighted mean grade and the performance on Mathematics is presented in Table \ref{T3}.

\begin{table}[H]
	\centering
	\caption{The Lift Function between the weighted mean grade and the performance on Mathematics. The numbers in parentheses represent the quantity of students in each category.}
	\label{T3}
	\begin{tabular}{ccccc}
		\hline
		Performance & \multicolumn{3}{c}{Weighted Mean Grade} & Relative \\ \cline{2-4}
		in Mathematics & Tertile 1 & Tertile 2 & Tertile 3 & Frequency \\ 
		\hline
		Tertile 1 & 1.3 (1,398) & 1.06 (1,111) & 0.631 (667) & 0.38 \\ 
		Tertile 2 & 0.935 (843) & 1.11 (972) & 0.956 (847) & 0.32 \\ 
		Tertile 3 & 0.689 (587) & 0.8 (661) & 1.51 (1,267) & 0.30 \\ 
		\hline
		Relative Frequency & 0.339 & 0.329 & 0.333 & 1 \\ 
		\hline
	\end{tabular}
\end{table}

The output of the algorithms provides relevant informations to the admission office of the University. Indeed, it is now known that the subjects that are most related to the performance on the undergraduate courses are Mathematics, Physics, Chemistry, Biology and Portuguese. Furthermore, in order to optimize the number of students that will succeed in the undergraduate courses, the office must select those that have high performance on Mathematics, as it \textit{lifts} by more than 50\% the probability of the student having also a high performance on the undergraduate course, i.e., students with high performance on Mathematics are prone to have high performance on the undergraduate course. Although the subjects that are most related to the performance on the courses are obtained from the classical feature selection algorithm, only the LLDS identifies what is the performance on the entrance exam that is most related to the success on the undergraduate course, that is high performance on Mathematics. Therefore, feature selection algorithms based on the LLDS provide more information than the classical feature selection algorithm, as they have a greater resolution and take into account the local relation between the variables.

\subsection{Congressional Voting Records dataset}

The Congressional Voting Records dataset consists of \textit{435} instances of \textit{16} Boolean features and a Boolean variable that indicates the party of the instance (democrat or republican). The features indicate how the instance voted (yes or no) in the year of 1984 about each one of \textit{16} matters, that are displayed in Table \ref{T4}. Algorithm \ref{A3} is applied to this dataset in order to determine what are the voting profiles that are most prone to be that of a republican and that of a democrat.

\begin{table}[H]
	\centering
	\caption{Features of the Congressional Voting Records dataset.}
	\label{T4}
	\begin{tabular}{ll}
		\hline
		ID & Matter (Feature) \\ 
		\hline
		HI & Handicapped infants \\ 
		WP & Water project cost sharing \\ 
		AB & Adoption of the budget resolution \\ 
		PF & Physician fee freeze \\ 
		SA & El Salvador aid \\ 
		RG & Religious groups in schools \\ 
		ST & Anti satellite test ban \\ 
		AN & Aid to nicaraguan contras \\ 
		MM & MX missile \\ 
		IM & Immigration \\ 
		SC & Synfuels corporation cutback \\ 
		ES & Education spending \\ 
		SR & Superfund right to sue \\ 
		CR & Crime \\ 
		DF & Duty Free exports \\ 
		EA & Export administration act South Africa \\ 
		\hline
	\end{tabular}
\end{table}

As the number of instances is relatively small, we perform Algorithm \ref{A3} under a restriction that avoids \textit{overfitting}. Indeed, if we apply the algorithm without the restriction, then the chosen profiles are those in which all the instances are of the same party. If there is only a couple of instances with some profile, and all of them are of the same party, then this profile is chosen as a \textit{prone} one for the party. However, we do not know if the profile is really prone, i.e., everybody with it is in fact of the same party, or if the fact of everybody with that profile being of the same party is just a sample deviation. In other words, without the restriction, the estimation error of the LF is too great as some profiles have low frequency in the sample and the feature selection algorithm \textit{overfits}.

Therefore, we restrict the search space to the profiles with a relative frequency in the sample of at least $0.15$. In other words, we select the profiles
\begin{equation*}
(\boldsymbol{\chi},\boldsymbol{x}_{opt}(y)) = \argmaxA_{\substack{\boldsymbol{\chi}^{*} \in \mathcal{P}(\boldsymbol{X}) \\ \boldsymbol{x}^{*} \in R_{\boldsymbol{\chi} ^{*}} \\ \mathbb{P}(\boldsymbol{\chi}^{*} = \boldsymbol{x}^{*}) > 0.15}} L_{\tiny (\boldsymbol{\chi^{*}},Y)}(\boldsymbol{x}^{*},y)
\end{equation*}
for $y \in \{democrat,republican\}$, in which $\mathbb{P}(\boldsymbol{\chi}^{*} = \boldsymbol{x}^{*}), \boldsymbol{\chi}^{*} \in \mathcal{P}(\boldsymbol{X}), \boldsymbol{x}^{*} \in R_{\boldsymbol{\chi}^{*}}$, is estimated by the relative frequency of the profile. The selected profiles, their LF value and the sample size considered are presented in Table \ref{T5}. At each iteration of the algorithm, only the instances that have no missing data in the features being considered are taken into account when calculating the LF, so that the sample size used at each iteration is not the same.

The profiles with maximum LF lifts by \textit{94\%} the probability of democrat and by around \textit{165\%} the probability of republican. This difference in the \textit{lift} is due to the fact that there are more democrats than republicans, so that the probability of democrat is greater and, therefore, can not be lifted as much as the probability of republican can. The profiles in Table \ref{T5} present a wide view of the voting profile of democrats and republicans, what allows an understanding of what differentiates a democrat from a republican regarding their vote.    

\begin{table}[ht]
	\small
	\centering
	\caption{Selected profiles obtained applying Algorithm \ref{A3} to the Congressional Voting Records dataset with the restriction that only the profiles with relative frequency greater than $0.15$ are considered. The instances with missing data were excluded at each iteration of the algorithm, i.e., $L$\tiny$_{(\boldsymbol{\chi{*}},Y)}$\normalsize$(\boldsymbol{x}^{*},y)$ is calculated using only the instances that have all the observations on the features $\boldsymbol{\chi}^{*}$.}
	\label{T5}	
	\begin{tabular}{clclc}
		\hline
		Party & Features ($\boldsymbol{\chi}$) & LF & Profile ($\boldsymbol{x}$) & Sample Size \\ 
		\hline
		\multirow{11}{*}{democrat} & (AB,PF,SA,RG,MM,ES,SR,EA) & 1.94 & (y,n,n,n,y,n,n,y) & 277 \\ 
		& (HI,AB,PF,SA,RG,MM,ES,SR,EA) & 1.94 & (y,y,n,n,n,y,n,n,y) & 275 \\ 
		& (AB,PF,RG,ST,MM,ES,SR,EA) & 1.94 & (y,n,n,y,y,n,n,y) & 279 \\ 
		& (HI,AB,PF,RG,ST,MM,ES,SR,EA) & 1.94 & (y,y,n,n,y,y,n,n,y) & 277 \\ 
		& (AB,PF,SA,RG,ST,MM,ES,SR,EA) & 1.94 & (y,n,n,n,y,y,n,n,y) & 276 \\ 
		& (HI,AB,PF,SA,RG,ST,MM,ES,SR,EA) & 1.94 & (y,y,n,n,n,y,y,n,n,y) & 274 \\ 
		& (AB,PF,SA,RG,MM,ES,SR,CR,EA) & 1.94 & (y,n,n,n,y,n,n,n,y) & 275 \\ 
		& (AB,PF,RG,ST,MM,ES,SR,CR,EA) & 1.94 & (y,n,n,y,y,n,n,n,y) & 276 \\ 
		& (AB,PF,SA,RG,ST,MM,ES,SR,CR,EA) & 1.94 & (y,n,n,n,y,y,n,n,n,y) & 274 \\ 
		& (AB,PF,RG,ST,MM,ES,SR,DF,EA) & 1.94 & (y,n,n,y,y,n,n,y,y) & 269 \\ 
		& (AB,PF,SA,RG,ST,MM,ES,SR,DF,EA) & 1.94 & (y,n,n,n,y,y,n,n,y,y) & 266 \\ 		
		\hline
		\multirow{16}{*}{republican} & (WP,PF,SC,ES,CR) & 2.65 & (n,y,n,y,y) & 342 \\
		& (AB,PF,AN,SC,CR,DF) & 2.64 & (n,y,n,n,y,n) & 369 \\
		& (PF,AN,IM,ES,CR,DF) & 2.64 & (y,n,y,y,y,n) & 361 \\ 
		& (PF,AN,SC,CR,DF) & 2.64 & (y,n,n,y,n) & 373 \\
		& (AB,PF,AN,SC,ES) & 2.63 & (n,y,n,n,y) & 376 \\ 
		& (HI,AB,PF,AN,SC,ES) & 2.63 & (n,n,y,n,n,y) & 373 \\ 
		& (AB,PF,AN,SC,ES,CR) & 2.63 & (n,y,n,n,y,y) & 368 \\ 
		& (HI,AB,PF,AN,SC,ES,CR) & 2.63 & (n,n,y,n,n,y,y) & 365 \\ 
		& (PF,AN,SC,DF) & 2.63 & (y,n,n,n) & 380 \\ 
		& (AB,PF,AN,SC,DF) & 2.63 & (n,y,n,n,n) & 376 \\ 
		& (PF,AN,IM,ES,DF) & 2.63 & (y,n,y,y,n) & 368 \\ 
		& (AB,PF,AN,SC,ES,DF) & 2.63 & (n,y,n,n,y,n) & 360 \\ 
		& (HI,AB,PF,AN,SC,CR,DF) & 2.63 & (n,n,y,n,n,y,n) & 365 \\ 
		& (PF,AN,SC,ES,CR,DF) & 2.63 & (y,n,n,y,y,n) & 356 \\ 
		& (AB,PF,AN,SC,ES,CR,DF) & 2.63 & (n,y,n,n,y,y,n) & 353 \\ 
		& (HI,AB,PF,AN,SC,ES,CR,DF) & 2.63 & (n,n,y,n,n,y,y,n) & 350 \\ 
		\hline
		\multicolumn{5}{l}{y = yes; n = no.}
	\end{tabular}
\end{table}

This application to the Congressional Voting Records dataset shed light on two interesting properties of the LLDS approach to feature selection in its higher resolution. First, this approach is indeed \textit{local}, as we are not interested in selecting the features that best classify the representatives accordingly to their party, but rather the voting \textit{profiles} that are most prone to be that of a democrat or republican. Secondly, the problem treated here is the opposite of a classification problem. Indeed, in the classification problem, we are interested in classifying a representative according to his party, given his voting profile. On the other hand, the problem treated here is the exact opposite: given a party, we want to know what are the profiles of the representatives that are most prone to be of that party. In other words, in the classification problem we want to determine the party given the voting profile, while on the LLDS problem we want to determine the voting profile given the party.

\subsection{Covertype dataset}

The Covertype dataset consists of \textit{581,012} instances (terrains) of \textit{54} features (10 continuous and 44 discrete) and a variable that indicates the cover type of the terrain (7 types). We apply Algorithms \ref{A1}, \ref{A2} and \ref{A3} to select features among the continuous ones that are displayed in Table \ref{T6}. The features are discretized in the same way they were in the performances dataset: by taking sample quantiles of the Mahalanobis distance between the features and zero. However, we now consider the quantiles $0.2,0.4,0.6$ and $0.8$ as cutting points, i.e., \textit{quintiles}, instead of tertiles.

\begin{table}[H]
	\centering
	\caption{Features of the Covertype dataset that are considered in this application.}
	\label{T6}
	\begin{tabular}{ll}
		\hline
		ID & Feature \\ 
		\hline
		E & Elevation \\ 
		A & Aspect \\ 
		S & Slope \\ 
		HH & Horizontal distance to hydrology \\
		HR & Horizontal distance to roadways \\ 
		HF & Horizontal distance to fire points \\
		H9 & Hillshade 9am \\ 
		HN & Hillshade Noon \\ 
		H3 & Hillshade 3pm \\ 
		VH & Vertical distance to hydrology \\ 
		\hline
	\end{tabular}
\end{table} 

Applying Algorithm \ref{A1} we select the features $\boldsymbol{\chi} = (\text{E,HH,HF})$, with a coefficient $\eta$\tiny$_{\boldsymbol{\chi}}$\normalsize$(Y \mid R_{\boldsymbol{\chi}}) = 0.307$ and the LF in Table \ref{T7}. We see that being in the first quintile of the selected features lifts classes 3, 4, 5 and 6; being in the second quintile lifts classes 2 and 5; being in the third quintile lifts class 2; being in the fourth quintile lifts class 1; and being in the fifth quintile lifts classes 1 and 7. From Table \ref{T7} we may interpret the relation between the selected features and the cover type. For example, we see that terrains with cover types 3, 4, 5 and 6 tend to have low joint values in the selected features, while terrains with cover 7 tend to have great joint values in them. This example shows how the proposed approach allows not only to select the features, but also understand why these features were selected, i.e., what is the relation between them and the cover type, by analysing the local dependence between the variables.

\begin{table}[H]
	\centering
	\caption{The Lift Function between the cover type of the terrain and the features Elevation, Horizontal distance to hydrology and Horizontal distance to fire points discretized by the sample quintiles of the Mahalanobis distance to zero. The numbers in parentheses represent the sample size of each category.}
	\label{T7}
	\resizebox{\linewidth}{!}{
		\begin{tabular}{ccccccccc}
			\hline
			\multirow{2}{*}{$(\text{E,HH,HF})$} & \multicolumn{7}{c}{Cover type} & Relative \\ \cline{2-8}
			& 1 & 2 & 3 & 4 & 5 & 6 & 7 & Frequency \\ 
			\hline
			Quintile 1 & 0.0766 (3,244) & 0.961 (54,473) & 4.94 (35,344) & 5 (2,747) & 1.78 (3,385) & 4.9 (17,010) & 0 (0) & 0.20 \\ 
			Quintile 2 & 0.444 (18,816) & 1.6 (90,872) & 0.0573 (410) & 0 (0) & 2.98 (5,663) & 0.103 (357) & 0.0205 (84) & 0.20 \\ 
			Quintile 3 & 0.949 (40,195) & 1.33 (75,562) & 0 (0) & 0 (0) & 0.234 (445) & 0 (0) & 0 (0) & 0.20 \\ 
			Quintile 4 & 1.66 (70,427) & 0.8 (45,314) & 0 (0) & 0 (0) & 0 (0) & 0 (0) & 0.112 (461) & 0.20 \\ 
			Quintile 5 & 1.87 (79,158) & 0.301 (17,080) & 0 (0) & 0 (0) & 0 (0) & 0 (0) & 4.87 (19,965) & 0.20 \\ 
			\hline
			Relative Frequency & 0.365 & 0.488 & 0.0615 & 0.00473 & 0.0163 & 0.0299 & 0.0353 & 1 \\ 
			\hline
		\end{tabular}}
	\end{table}
	
	Applying Algorithm \ref{A2} to this dataset we obtain the windows displayed in Table \ref{T8}. We see that the window that seems to most influence the cover type is the first and fifth quintile of the features Elevation and Horizontal distance to hydrology. Indeed, all the top ten windows contain those two features, and either their first or fifth quintile. As we can see in Table \ref{T7}, the influence of the fifth quintile of $\boldsymbol{\chi} = (\text{E,HH,HF})$, the top window, is given by the fact that no terrain of the types 3, 4, 5 and 6 is in this quintile. Note that, again, our approach allows a better interpretation of the selected features by the analysis of the local dependence between the features and the cover type.
	
	\begin{table}[H]
		\centering
		\caption{Features selected applying Algorithm \ref{A2} to the Covertype dataset.}
		\label{T8}
		\begin{tabular}{llc}
			\hline
			Features ($\boldsymbol{\chi}$) & Window ($W$) & $\eta$\tiny$_{\boldsymbol{\chi}}$\normalsize$(Y \mid W)$ \\ 
			\hline
			(E,HH,HF) & Quintile 5 & 0.38 \\ 
			(E,A,HH,HF) & Quintile 5 & 0.38 \\ 
			(E,HH) & Quintile 5 & 0.37 \\ 
			(E,A,HH) & Quintile 5 & 0.37 \\ 
			(E,HH,VH,HF) & Quintile 5 & 0.37 \\ 
			(E,A,HH,VH,HF) & Quintile 5 & 0.36 \\ 
			(E,HH,HF) & Quintiles 1  \& 5 & 0.36 \\ 
			(E,A,HH,VH) & Quintile 5 & 0.36 \\ 
			(E,HH,VH) & Quintile 5 & 0.36 \\ 
			(E,HH) & Quintiles 1 \& 5 & 0.36 \\ 
			\hline
		\end{tabular}
	\end{table}
	
	Finally, applying Algorithm \ref{A3} we choose the profiles displayed in Table \ref{T9} for $y \in \{1,2,3,4,5,6,7\}$. We see, for example, that the profile most prone to be of type 1 is $(\text{E,HH,HF}) = \text{Quintile 5}$ and of type 3 is $(\text{E,HH,HR,HF}) = \text{Quintile 1}$. Note that it does not mean that most of the terrains with these profiles are of type 1 and 3, but rather that the probability of a terrain with these profiles being of types 1 and 3, respectively, is \textit{87\%} and \textit{396\%} greater than the probability of a terrain for which we do not know the profile. Therefore, we see again the difference between the LLDS approach and the classification problem. In the LLDS approach, given a profile, we are interested in determining the type of which the conditional probability given the profile is greater than the marginal probability, while in the classification problem, given a profile, we are interested in determining the type for which the conditional probability given the profile is the greatest.
	
	As an example, suppose the joint distribution that generated the LF of Table \ref{T7} and the profile Quintile 1. We have that the maximum conditional probability given this profile is the probability of type 2 ($54,473/116,203 = 0.47$), while the maximum lift is that of type $4$,although its conditional probability is only $2,747/116,203 = 0.02$. However, the conditional probability of type $4$ given the profile, even though absolutely small, is relatively great: it is 5 times the marginal probability $0.004$. Therefore, on the one hand, if there is a new terrain whose profile is $(\text{E,HH,HF}) = \text{Quintile 1}$, we classify it as being of type 2. On the other hand, if we want to sample terrains from a population and are interested in maximizing the number of terrains of type 4, we may sample from the population with profile $(\text{E,HH,HF}) = \text{Quintile 1}$ instead of the whole population, expecting to sample four times more terrains of type 4.
	
	\begin{table}[H]
		\centering
		\caption{Profiles selected applying Algorithm \ref{A3} to the Covertype dataset for $y \in \{1,2,3,4,5,6,7\}$.}
		\label{T9}
		\begin{tabular}{clcc}
			\hline
			Cover type & Features & LF Maximum & Profile \\ 
			\hline
			1 & (E,HH,HF) & 1.87 & Quintile 5 \\ 
			2 & (E,HH,HR) & 1.63 & Quintile 2 \\ 
			3 & (E,HH,HR,HF) & 4.96 & Quintile 1 \\ 
			4 & (E,HH,HF)$^{1}$ & 5 & Quintile 1 \\ 
			5 & (E,HR,HF) & 3.31 & Quintile 2 \\ 
			6 & (E,HH,HF) & 4.90 & Quintile 1 \\ 
			7 & (E,HH) & 4.89 & Quintile 5 \\ 
			\hline
			\multicolumn{4}{l}{\footnotesize $^{1}$Among other profiles.} \\
		\end{tabular}
	\end{table}
	
	\section{Final Remarks}
	
	The feature selection algorithms based on the LLDS extend the classical approach to feature selection  to a higher resolution one, as it takes into account the local dependence between the features and the variable of interest. Indeed, classical feature selection may be performed by walking through a tree in which each node is a vector of features, i.e., a BLFS, while feature selection based on the LLDS is established by walking through an extended tree, i.e., a CBLOP, in which inside each node there is another tree, that represents the windows of the variables, as displayed in the example in Figure \ref{FS}. Therefore, feature selection based on the LLDS increases the reach of feature selection algorithms to a new variety of applications.
	
	The LLDS may treat a problem that is the opposite of that of classification, i.e., when we are interested in, given a class $y$, finding the profile $\boldsymbol{x}$ of which we may sample from its population in order to maximize the number of instances of class $y$. Indeed, in the classification problem we want to do the exact opposite: classify a instance with known profile $\boldsymbol{x}$ into a class of $Y$. Therefore, although LLDS tools may also be applied to the classification problem (as they are in the literature), they are of great importance in problems that we may call the \textit{reverse engineering} of the classification one. Thus, our approach broadens the application of features selection algorithms to a new set o problems by the extension of their search spaces from BLFs to CBLOPs.
	
	The algorithm proposed in this paper may be optimized in order to not walk through the entire CBLOP, as its size increases exponentially with the number of features, so that the algorithm may not be computable for a great number of features. Moreover, the algorithms may be subjected to \textit{overfitting} if the sample size is relatively small, so that their search space may be restricted. The methods of \cite{guyon2003,kohavi1997,lal2006,john1994,hall2000,das2001,yu2003,marill1963,whitney1971,pudil1994,somol1999,somol2006,nakariyakul2009,narendra1977,u-curve1,u-curve2,u-curve3}, for example, may be adapted to the multi-resolution algorithms in order to optimize them. Furthermore, the properties of the $\eta$ coefficients and the LF must be studied in a theoretical framework, in order to establish their variances, sample distributions and develop statistical methods to estimate and test hypothesis about them. 
	
	The LLDS adapts classical measures, such as the MI and the Kullback-Leibler Divergence, into coherent dependence coefficients that assess the dependence between random variables in multiple resolutions, presenting a wide view of the dependence between the variables. As it does not make any assumption about the dependence kind, the LLDS measures the raw dependence between the variables and, therefore, may be relevant for numerous purposes, being feature selection just one of them. We believe that the algorithms proposed in this paper, and the LLDS in general, bring advances to the state-of-art in dependence measuring and feature selection, and may be useful in various frameworks.

\vspace{6pt} 

\section*{Supplementary Material}
The following are available online at www.mdpi.com/link: an \textbf{R} \cite{R} package called \textit{localift} that performs the algorithms proposed by this paper; an \textbf{R} object that contains the results of the algorithms used to analyse the performances dataset; and an \textbf{R} code that apply the algorithms to the Congressional Voting Records and Covertype datasets.

\section*{Acknowledgements}
We would like to thank A. C. Hernandes who kindly provided the performances dataset used in the application section. The Covertype dataset is Copyrighted 1998 by Jock A. Blackard and Colorado State University.

	
	\bibliographystyle{plain}
	\bibliography{Ref}

\end{document}